 \documentstyle[prl,aps]{revtex}   
 \def\ket{\!>\,} \def\ack{\,|\,}

\begin{document}
\draft


  \twocolumn[\hsize\textwidth\columnwidth\hsize  
  \csname @twocolumnfalse\endcsname              

\title{
Multi-Phonon $\gamma$-Vibrational Bands  
and the Triaxial Projected Shell Model
}

\author{Yang Sun$^{(1,2)}$, Kenji Hara$^{(3) \dag}$, 
Javid A. Sheikh$^{(4)}$, 
Jorge G. Hirsch$^{(5)}$,   Victor Vel\'azquez$^{(6)}$ and Mike
Guidry$^{(1)}$}

\address{
$^{(1)}$Department of Physics and Astronomy, University of Tennessee,
Knoxville, TN 37996, U.S.A.\\
$^{(2)}$Department of Physics, Xuzhou Normal University,
Xuzhou, Jiangsu 221009, P.R. China\\
$^{(3)}$Physik-Department, Technische Universit\"at M\"unchen,
D-85747 Garching bei M\"unchen, Germany\\
$^{(4)}$Tata Institute of Fundamental
Research, Colaba, Bombay -- 400 005, India\\
$^{(5)}$ Instituto de Ciencias Nucleares, UNAM, Circuito Exterior
C.U., A.P. 70-543,   04510 M\'exico D.F., M\'exico\\
$^{(6)}$ Departamento de F\'{\i}sica, Centro de Investigaci\'on y
Estudios Avanzados del IPN,  
A.P. 14-740, 07000 M\'exico D.F., M\'exico
}

\maketitle

\begin{abstract}
We present a fully quantum-mechanical, microscopic, unified  treatment of 
ground-state band and multi-phonon $\gamma$-vibrational bands using
shell model diagonalization with
the triaxial projected shell model.
The results agree very well with data on the g-
and $\gamma$-band spectra in $^{156-170}$Er, as well as with recently measured
$4^+$ 2-phonon $\gamma$-bandhead energies in $^{166}$Er and
$^{168}$Er. Multi-phonon $\gamma$-excitation energies are predicted.
\end{abstract}


  ]  


The atomic nucleus is a many-body system with pronounced shell effects
that can have intrinsic deformation. In addition, it can,
according to the semi-classical collective model, undergo dynamical
oscillations around the equilibrium shape, resulting in various
low-lying collective excitations.
Ellipsoidal oscillation of the shape 
is commonly termed a
$\gamma$-vibration \cite{BM75}.

Thanks to advances in high-resolution $\gamma$-ray detectors,
high quality measurements not only of high-spin states but also of
low-spin states are now commonly available.  As a consequence
long-sought multi-phonon $\gamma$-vibrational states
have been discovered in a series of experiments over the last decade
\cite{Bo91,Ko93,Fa96,Ga97,Co97,Ha98}. However,
the status of unified theoretical descriptions for
ground-state band (g-band) and multi-phonon
$\gamma$-vibrational bands ($\gamma$-band) 
is not so satisfactory. In the
present work, we attempt a consistent description of
these low-lying bands using an approach based on the Projected Shell
Model (PSM) \cite{review}.

In its original form, the PSM uses an axially deformed basis. The
shell model diagonalization is carried out within the space spanned by
the angular momentum 
projected quasiparticle (qp) vacuum, 2- and 4-qp states. In this
sense, the PSM is a Tamm--Dancoff approach and one expects that the
collectivity of low-lying states may be strongly affected by mixing many
2- and 4-qp states. Indeed, a multi-qp admixture can cause significant
effects in band crossing regions \cite{review}.

However, in the low-spin region before any band crossings
($I \le 10$), the admixture is very weak and the calculated 
g-band always exhibits the characteristics of an axially
symmetric rotor. For example, the ordinary PSM fails to describe the
steep increase of moment of inertia at low spins in transitional nuclei
\cite{review}. Quite recently, the restriction to an axially deformed
basis in the PSM was removed by two of the present authors (JAS and KH).
It was shown that the observed steep increase of moment of inertia for
transitional nuclei can be well described if one introduces triaxiality
in the deformed basis and performs 3-dimensional angular momentum
projection \cite{SH99}. This approach is called the Triaxial Projected
Shell Model (TPSM).

Another important issue is whether the PSM can 
describe bands built on collective vibrational states.
The usual treatment of the $\gamma$-band based on the
Tamm--Dancoff or on the Random Phase Approximation assumes different
coupling constants for the $\mu=0$ and $\mu=\pm2$ parts of the QQ-force,
with the former related to the mean-field deformation and the latter
adjusted to the $\gamma$-bandhead energy. In the PSM, as in the ordinary
shell model, such an adjustment is not permitted  because the Hamiltonian
must be rotation-invariant and thus these two coupling constants must
be equal: One cannot simply fit the theoretical
$\gamma$-bandhead to the experimental one by modifying the QQ-force in
that manner. 

On the other hand, one might hope that inclusion of many
2-qp states could introduce a collective contribution that would produce
the desired low-lying $\gamma$-state. But such attempts have failed. Because of
the large pairing gap, the energy of the lowest 2-qp state is above 1.5
MeV and is much higher than the actual $\gamma$-bandhead energy, which 
typically
lies between 0.5 and 1 MeV in rare-earth nuclei. The QQ-force is
too weak to lower the theoretical $\gamma$-band energy by such a
large amount in a limited basis. Calculations including 
about one thousand 2- and 4-qp
states do not lead to low-lying excited states that
look like the experimental $\gamma$-band \cite{Sun}. One therefore has
to conclude that it is not practical to describe the
$\gamma$-vibrational state in terms of multi-qp states in the framework
of the axial PSM \cite{Vel98}.

In the present paper, the TPSM extension of the PSM and the computer
code developed in \cite{SH99} are used to study 
multi-phonon $\gamma$-bands.  (Although
the present theory is not based on a vibrational phonon
excitation mechanism as in other models \cite{So90}, 
we shall use the conventional vibrational terminology in our discussion.)
We shall show the following: (1)~For well deformed nuclei, introduction
of triaxiality in the basis does not destroy the good agreement for the
g-bands obtained previously in the axial PSM calculations (for example,
those presented in Ref.\ \cite{review}). (2)~However, it produces new
excited states ($\gamma$-bands) at the correct energies
that do not occur in the axial PSM.
(3)~For transitional nuclei,
use of a basis of fixed triaxiality improves the g-band moments of inertia, 
as already shown
in \cite{SH99}, and at the same time produces realistic $\gamma$-bands.
(4) By a single diagonalization of
the Hamiltonian (with the same parameters in the deformed basis), we
obtain not only the g- and $\gamma$-band, but also higher excited bands that can
be identified as the multi-phonon $\gamma$-bands and these
compare very well with recently measured $4^+$ 2-phonon $\gamma$-bands.
(5) Finally, we make predictions for the 2- and 3-phonon $\gamma$-band
(referred to as $2\gamma$- and $3\gamma$-band hereafter) of those Er
isotopes treated here for which no measurement has yet been reported.

Since 
an extensive review of the PSM exists (see Ref.\ \cite{review} and references
cited therein), we shall describe the model only briefly. The PSM (TPSM)
closely follows the shell model philosophy and is, in fact, a shell
model truncated in a deformed basis. One uses a Nilsson potential having
axial (triaxial) deformation to generate the deformed single-particle
states. The Nilsson spin--orbit force parameters $\kappa$ and $\mu$ 
are essential in reproducing correct shell fillings. For
rare-earth nuclei, we use the early compilation of Nilsson {\it et
al} \cite{NKM} without modification. For the axial deformation parameter
$\epsilon$ in the Nilsson model, we take the values given in Ref.\
\cite{BFM86}. Thus, for the TPSM, the triaxial deformation $\epsilon'$
is the single adjustable parameter. The static pairing correlations are
treated by the usual BCS approximation to establish the Nilsson+BCS
basis. The 3-dimensional angular momentum projection is then carried out
on the Nilsson+BCS qp-states to obtain the many-body basis, and the
Hamiltonian is diagonalized in this projected basis.

In the present work, we consider only low-spin states 
where no band crossing with any multi-qp band occurs in the yrast region. 
Thus, the many-body
basis may be restricted to the projected triaxial qp vacuum state:
\begin{equation}
\left\{\hat P^I_{MK}\ack\Phi\ket,~0 \le K \le I \right\},
\label{basis}
\end{equation}
where $\ack\Phi\ket$ represents the triaxial qp vacuum state. This is
the simplest possible configuration space for an even--even nucleus. Note
that only one state is possible for spin $I=0$ (the ground state). Thus,
multi-qp components have to be taken into account if one wants to
describe $I=0$ excited states (see further discussion below). The
diagonalization is performed over a chain of Er isotopes up to spin
$I=10$. 

As in the usual PSM calculations, we use the Hamiltonian
\cite{review}
\begin{equation}
\hat H = \hat H_0 - {1 \over 2} \chi \sum_\mu \hat Q^\dagger_\mu
\hat Q^{}_\mu - G_M \hat P^\dagger \hat P - G_Q \sum_\mu \hat
P^\dagger_\mu\hat P^{}_\mu,
\label{hamham}
\end{equation}
so that the corresponding Nilsson Hamiltonian (with triaxiality) is
given by
\begin{equation}
\hat H_N = \hat H_0 - {2 \over 3}\hbar\omega\left\{\epsilon\hat Q_0
+\epsilon'{{\hat Q_{+2}+\hat Q_{-2}}\over\sqrt{2}}\right\}.
\label{nilsson}
\end{equation}
Here $\hat H_0$ is the spherical single-particle Hamiltonian, which
contains a proper spin--orbit force as mentioned before,
while the interaction strengths are taken as follows. The QQ-force
strength $\chi$ is adjusted such that the physical  quadrupole
deformation $\epsilon$ is obtained as a result of the self-consistent
mean-field (HFB) calculation \cite{review}. The monopole pairing
strength $G_M$ is of the standard form $G_M = \left[21.24
\mp13.86(N-Z)/A\right]/A$, with ``$-$" for neutrons and ``$+$" for
protons, which approximately reproduces the observed odd--even mass
differences in this mass region. This choice of $G_M$ is appropriate for
the single-particle space employed in the PSM, where three major shells
are used for each type of nucleons ($N=4,5,6$ for neutrons and $N=3,4,5$
for protons). The quadrupole pairing strength $G_Q$ is assumed to be
proportional to $G_M$, the proportionality constant being fixed as usual
to be in the range 0.16 -- 0.18. These interaction strengths are
consistent with those used previously for the same mass region
\cite{review,SH99}.

Let us first consider a well-deformed nucleus $^{168}$Er,
which is generally considered to be axially symmetric. In fact,
previous (axial) PSM calculation for this nucleus gave an excellent  
description of the yrast band up to a very high spin \cite{review}. 
Fig.\ 1a shows the calculated energies as functions of the triaxiality
parameter $\epsilon'$ for angular momenta up to $I=10$. In addition to
the usual g-band with spins $I=0,2,4,\cdots$, a new set of rotational
states with spins $I=2,3,4,\cdots$ appears. This figure looks similar to
the one shown by Davydov and Filippov \cite{DF58}, but  now obtained in
terms of a fully microscopic theory. Unlike the irrotational flow model,
the PSM spectrum depends not only on the deformation parameters but
also on the shell filling of the nucleus in question. We see that, for
the g-band of $^{168}$Er, the energies as functions of triaxiality are
nearly flat and their values remain close to those at zero triaxiality.
Thus, the triaxial basis has no significant effect on the g-band for a
well-deformed nucleus and does not destroy the good g-band result
obtained with an axially deformed basis. 

However, it has a drastic
effect on new excited bands (second and higher excited bands are
not shown in the figure). Their excitation energies are indeed very high
for axial symmetry,  but come down quickly as the triaxiality in
the basis increases. At
$\epsilon'=0.13$, the first excited band reproduces the observed
$\gamma$-band in $^{168}$Er (while preserving the good g-band
agreement). It should be noted that the excited bands studied 
in this paper are obtained by introducing $\gamma$-degree of freedom 
in the basis
(quasiparticle vacuum). 
They are collective excitations, but not quasiparticle excitations.
We may thus identify the first excited band as the
$\gamma$-band, the second excited band as the
2$\gamma$-band, the third excited band as the
3$\gamma$-band, etc.

The above results can be understood by studying the $K$-mixing
coefficients for each projected $K$-state (see Eq. (\ref{basis})) in the
total wavefunctions. 
It is found that for this well-deformed, axially symmetric nucleus, 
$K$-mixing is negligibly small. 
States in the g-band are essentially the projected $K=0$ state for any
$\epsilon'$.
That is why the basis triaxiality does not destroy the result
obtained with an axially deformed basis.
The excited bands are also built by rather pure projected $K$-states. 
For example, the first excited band with the bandhead spin $I=2$ 
is manily the projected $K=2$ state 
and the second excited band with the bandhead spin $I=4$ 
is the projected $K=4$ state. 
A small amount of $K$-mixing can be seen only  
for states with higher total spin 
if triaxiality in
the basis is sufficientlly large. 

Fig. \ 1b illustrates another example, 
the transitional nucleus 
$^{156}$Er. 
We see that the energies for the g-band are no longer
constant, but clearly vary as functions of triaxiality.
This feature is expected for a $\gamma$-soft nucleus. 
For the excited bands, triaxiality in the basis has a similar
effect as we have seen for well-deformed nucleus: it drastically
lowers their energies to those of the observed $\gamma$-band. 

A rather different picture of $K$-mixing is observed for this
$\gamma$-soft nucleus.
The states are no longer pure projected $K$-states, but highly mixed. 
For example, the two $I=2$ states (the one in the g-band and the other one
being the bandhead of the first excited band (the $\gamma$-band) 
are mixed from the projected $K=0$ and $K=2$ states. 
At $\epsilon'=0.13$, the $I=2$ state of the g-band is contaminated by 
the projected $K=2$ state with a weight of about 1/4, 
and the $I=2$ state of the first excited band contains the projected 
$K=0$ state with a weight of about 1/4.
Stronger $K$-mixing is seen for states with higher total spin and
larger basis triaxiality. 

Fig.\ 2 presents results for a chain of Er isotopes with neutron numbers
from $N=88$ to 102. This covers both transitional ($N \approx 90$) and
well-deformed ($N \approx 98$) nuclei. The theoretical results are
compared with available data for both g- and $\gamma$-band up to $I=10$.
The axial and triaxial deformation parameters used in the present
calculations are listed in Table I.
The triaxial parameter $\epsilon'=0.13$
giving the correct position of the $\gamma$-band
for $^{168}$Er and $^{156}$Er corresponds to $\gamma=
25.5^\circ$ in terms of the usual gamma parameter, if one uses as a very rough
estimate $\gamma \approx \tan^{-1} ({\epsilon'/\epsilon})$.

A microscopic description of transitional nuclei has always been 
challenging. The nuclei discussed here with neutron number around 90
have g-bands that are quasi-rotational but with considerable
vibrational character. The ground-state energy surface of a transitional
nucleus was shown to have a shallow minimum at a finite
$\gamma$-deformation in HFB calculations \cite{KB68}. It has
been demonstrated that such a shallow minimum becomes a prominent
minimum when projected onto spin $I=0$ \cite{HHR82}. The necessity of
introducing triaxiality in the PSM basis to describe the observed g-band
moment of inertia in transitional nuclei was demonstrated in Ref.\
\cite{SH99}. We now see that, with the same triaxiality, the first
excited TPSM band reproduces also the observed $\gamma$-band. By adjusting a
single parameter $\epsilon'$ in the TPSM, the spectra of both the g- and
$\gamma$-band are described simultaneously and consistently by
the Hamiltonian (\ref{hamham}) diagonalized within the Hilbert space
(\ref{basis}).
 
Next, let us turn to a discussion of multi-phonon $\gamma$-bands. In
Fig.\ 3, we plot all the states for spins $I \le 10$ obtained after
diagonalization within our projected triaxial basis for two nuclei in
which a 2$\gamma$-band has been reported.
For $^{168}$Er, the second excited theoretical band agrees beautifully with
the new $4^+$ $2\gamma$-band reported in Ref.\ \cite{Ha98}.
For $^{166}$Er, the observed $4^+$ $2\gamma$-bandhead \cite{Fa96} is
also well reproduced. Since our theory agrees very well with the g-band
and the (1-phonon) $\gamma$-band observed in these nuclei, the present
results support strongly the interpretation of these data as 
2$\gamma$-bands. 

To our knowledge, no 
3$\gamma$-band has yet been seen experimentally. According to our
calculations, they should appear between 3 and 3.6 MeV. In Table I, we
list the theoretical values for the $2^+$ $\gamma$-, $4^+$ $2\gamma$-
and $6^+$ $3\gamma$-bandhead energies. As Table I shows, the predicted
$\gamma$-vibrational spectra are quite anharmonic. 
Anharmonic $\gamma$-vibrations have been discussed by several
authors \cite{DH82,MM86,MN93}.
This anharmonicity is
a straightforward consequence of the present microscopic theory. 
This may be contrasted with earlier models that found it
necessary to introduce explicit anharmonicities to reproduce the
$\gamma$-band spacings \cite{DH82}.

Finally, we mention briefly the $0^+$ excited states.
Unlike the usual collective models based on phonon excitations,
a $0^+$ collective excited state does not exist in the present
calculation. Excited $0^+$ states can occur if we include multi-qp
states on top of the present vacuum configuration. However, since the
states constructed in this way are mainly qp in character,
the collectivity of such a $0^+$ excited state is generally expected to be 
much weaker than that of a 2-phonon $\gamma$-state, 
which should have a large E2-decay
probability to a 1-phonon $\gamma$-state. Furthermore, such states should 
depend strongly on the shell fillings. Therefore, the nature of the $4^+$ 
2-phonon excited state 
is kinematical while the $0^+$ 2-phonon excited 
state is dynamical. There
has been one experiment reporting a $0^+$ excited state
in $^{166}$Er \cite{Ga97}; the  measured B(E2:$0^+\!\!
\rightarrow \!\!2^+_\gamma$) is enhanced, suggesting that this $0^+$
excited state is of  2-phonon nature. At present, this is a single
observed example of the $0^+$ 2-phonon excited state. Whether this
observation can be reproduced by the TPSM with inclusion of qp states remains
to be seen.

To summarize, we have applied the Triaxial Projected Shell Model
to some Er isotopes to
investigate multi-phonon $\gamma$ vibrational bands. 
The shell model diagonalization is not carried out in a spherical
basis as for a conventional shell model, but in a deformed basis
with triaxiality. It is found
that this simultaneously improves the description of the g-bands in
transitional nuclei and leads to a consistent description of 
multi-phonon $\gamma$-bands in both transitional
and well-deformed nuclei. The newly observed $4^+$
$2\gamma$-bands are reproduced by the same calculation, thus supporting their
experimental assignment, and the  bandhead energies of as yet unobserved
$6^+$ $3\gamma$-bands  are predicted.

Thus, our unified view of the g- and
multi-phonon $\gamma$-bands agrees surprisingly well with the
existing data, even though  we have used the simplest possible configuration
space. The origin of the $\gamma$-bands discussed in the present paper
is kinematical rather than
dynamical, indicating a microscopic connection between the
$\gamma$-excited states and the nuclear ground state properties. We
are presently investigating various intra- and inter-band B(E2)-values to
test the theory further. These results will be discussed in terms of
$K$-mixing and reported in a longer paper. 

Dr. Kenji Hara$^{\dag}$ worked on the present paper until his last day. This
Letter is dedicated to the memory of his lifetime contributions to the 
Projected Shell Model. 
This work was supported in part by Conacyt (Mexico).

$\dag$ deceased.

\baselineskip = 16pt
\bibliographystyle{unsrt}

\begin{figure}
\caption{ 
Calculated energies (solid lines) of the g- and $\gamma$-band 
in (a) $^{168}$Er and (b) $^{156}$Er
as functions of triaxiality parameter $\epsilon'$
for angular momenta up to $I = 10$. 
The experimental g-band (open circles) and $\gamma$-band (open 
triangles) are best reproduced by the TPSM at $\epsilon' = 0.13$. 
Data are taken from \protect\cite{ISO}. 
}
\label{figure.1}
\end{figure}
\begin{figure}
\caption{ 
Comparison of calculated energies for the g-band (open circles) and
$\gamma$-band (open rectangles) with the available experimental data
for $^{156-170}$Er \protect\cite{ISO} (filled diamonds for
g-bands and filled triangles for $\gamma$-bands).
}
\label{figure.2}
\end{figure}
\begin{figure}
\caption{ 
The spectra up to $I=10$ for $^{166,168}$Er. Theoretical results are
compared with the available experimental data for the g-band and
$\gamma$-band \protect\cite{ISO}, as well as the $4^+$ $2\gamma$-band in
$^{166}$Er \protect\cite{Fa96} and $^{168}$Er \protect\cite{Ha98}.
}
\label{figure.3}
\end{figure}

\begin{table}[h]
\begin{center}
\caption{Axial and triaxial quadrupole deformation parameters $\epsilon$
and $\epsilon'$ employed in the TPSM calculation for Er isotopes. 
$E_{2\gamma}(4^+)$ and $E_{3\gamma}(6^+)$ are predicted energies for the
$\gamma$-, $2\gamma$- and $3\gamma$-bandhead in units of MeV. Note their
anharmonicity.}
\begin{tabular}{c|c|c|c|c|c}
 A & $\epsilon$ & $\epsilon'$ & $E_{\gamma} (2^+)$ &
 $E_{2\gamma} (4^+)$ & $E_{3\gamma} (6^+)$ \\ \hline
156 & 0.200 & 0.13 & 0.824 & 2.090 & 3.567 \\ \hline
158 & 0.215 & 0.14 & 0.650 & 1.774 & 3.116 \\ \hline
160 & 0.230 & 0.14 & 0.658 & 1.816 & 3.180 \\ \hline
162 & 0.245 & 0.13 & 0.808 & 2.130 & 3.636 \\ \hline
164 & 0.258 & 0.14 & 0.743 & 2.084 & 3.655 \\ \hline
166 & 0.267 & 0.14 & 0.744 & 2.109 & 3.692 \\ \hline
168 & 0.273 & 0.13 & 0.778 & 2.085 & 3.478 \\ \hline
170 & 0.276 & 0.11 & 0.967 & 2.276 & 3.470
\end{tabular}
\end{center}
\end{table}

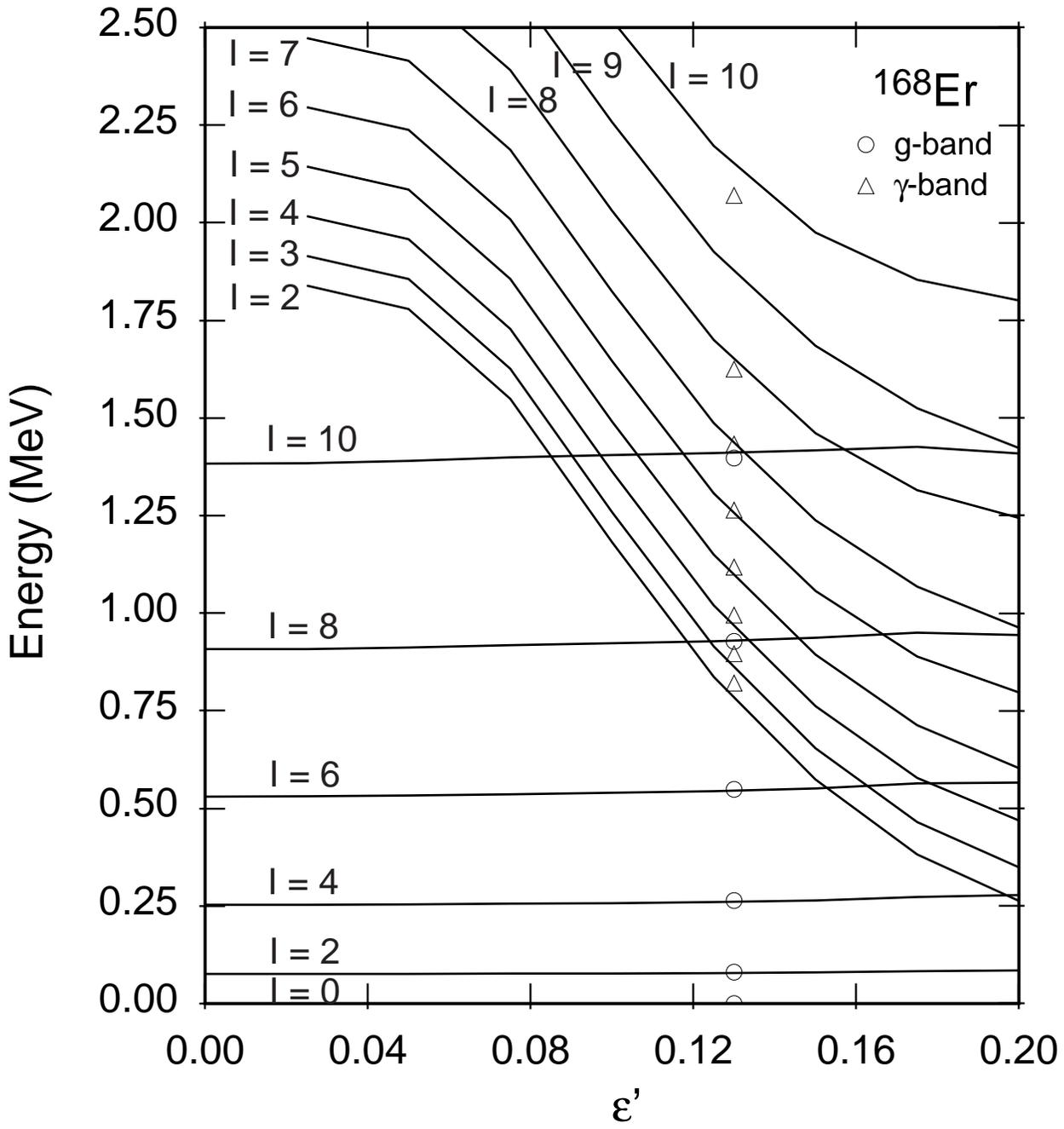
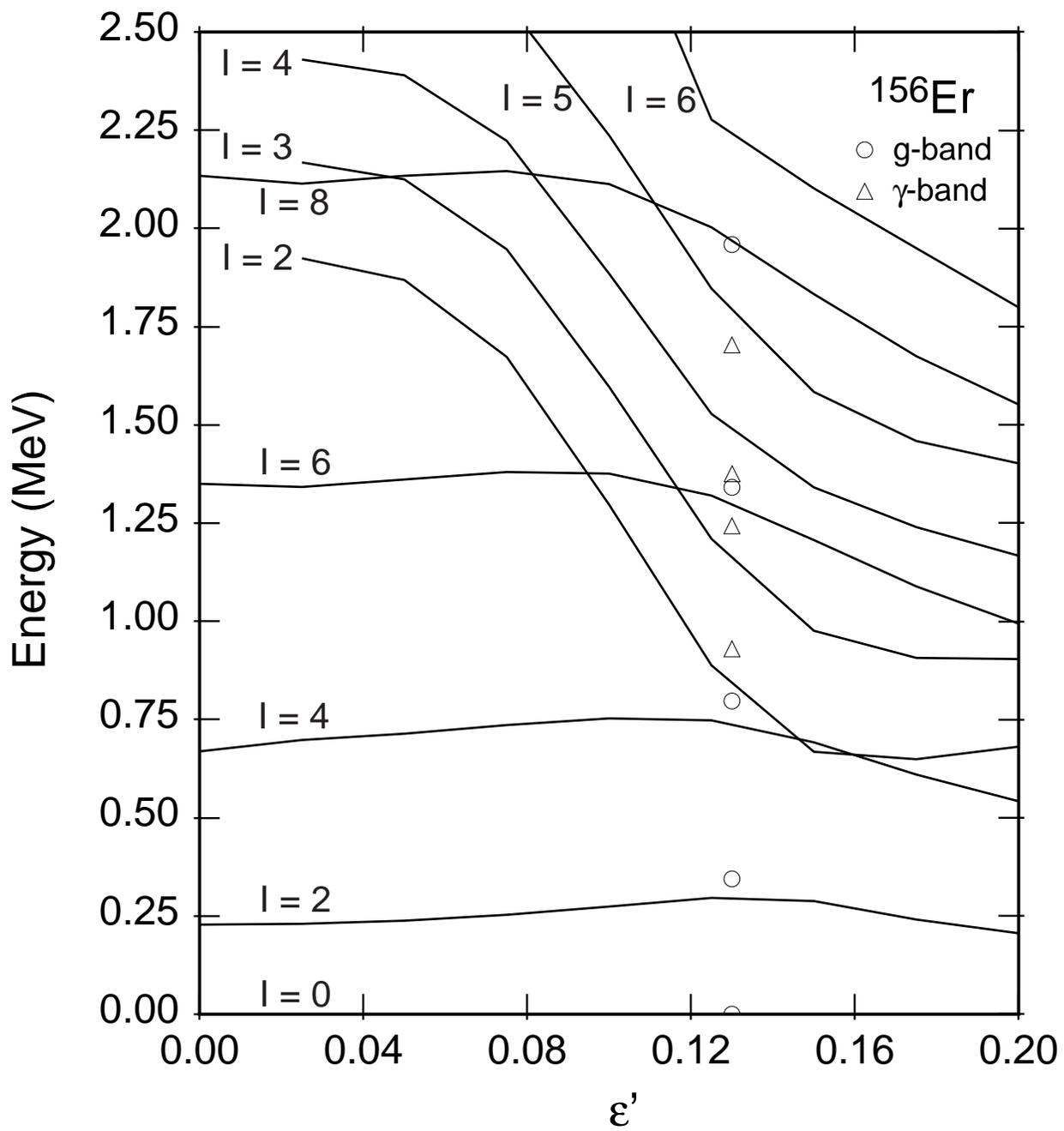
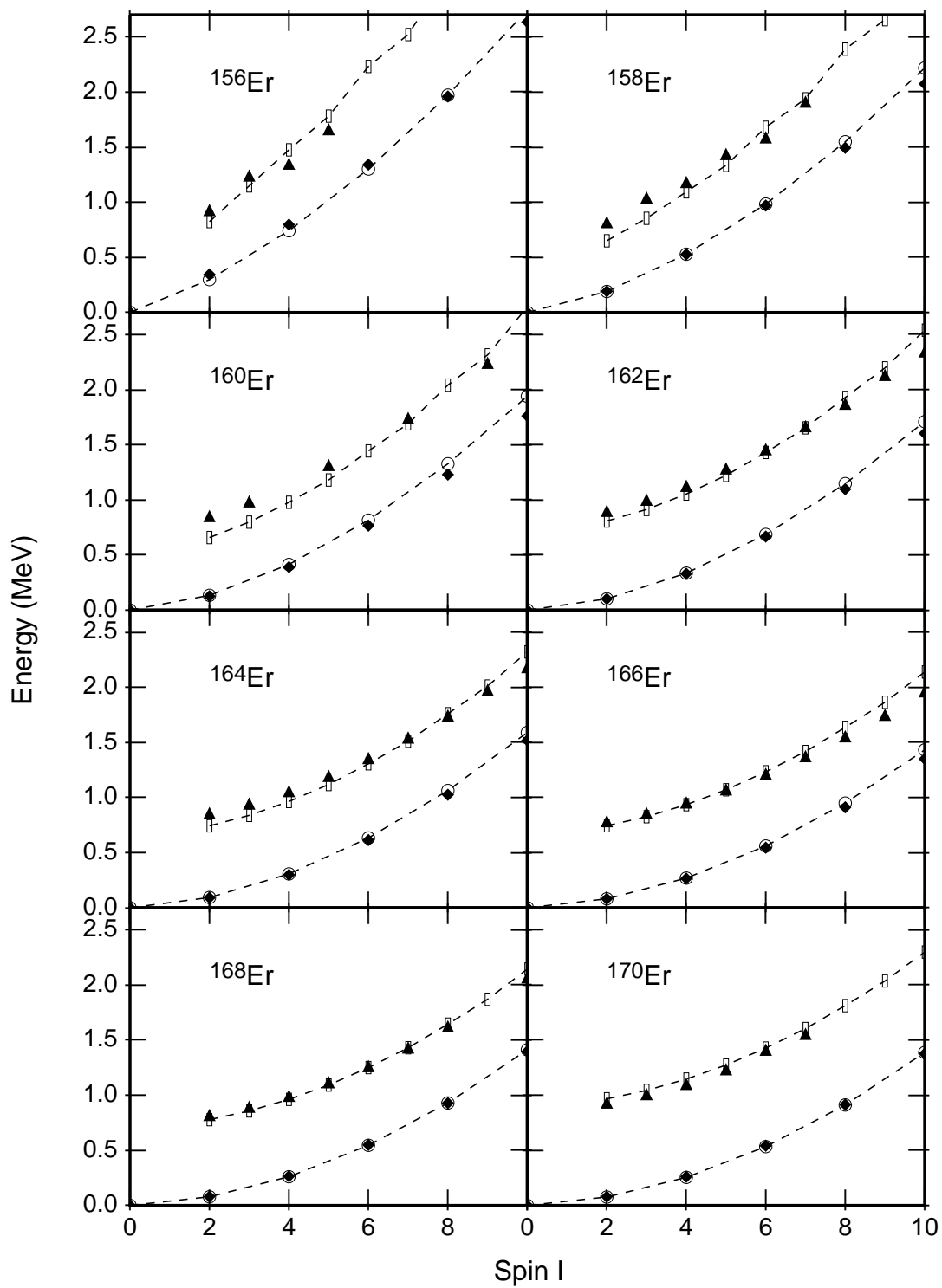
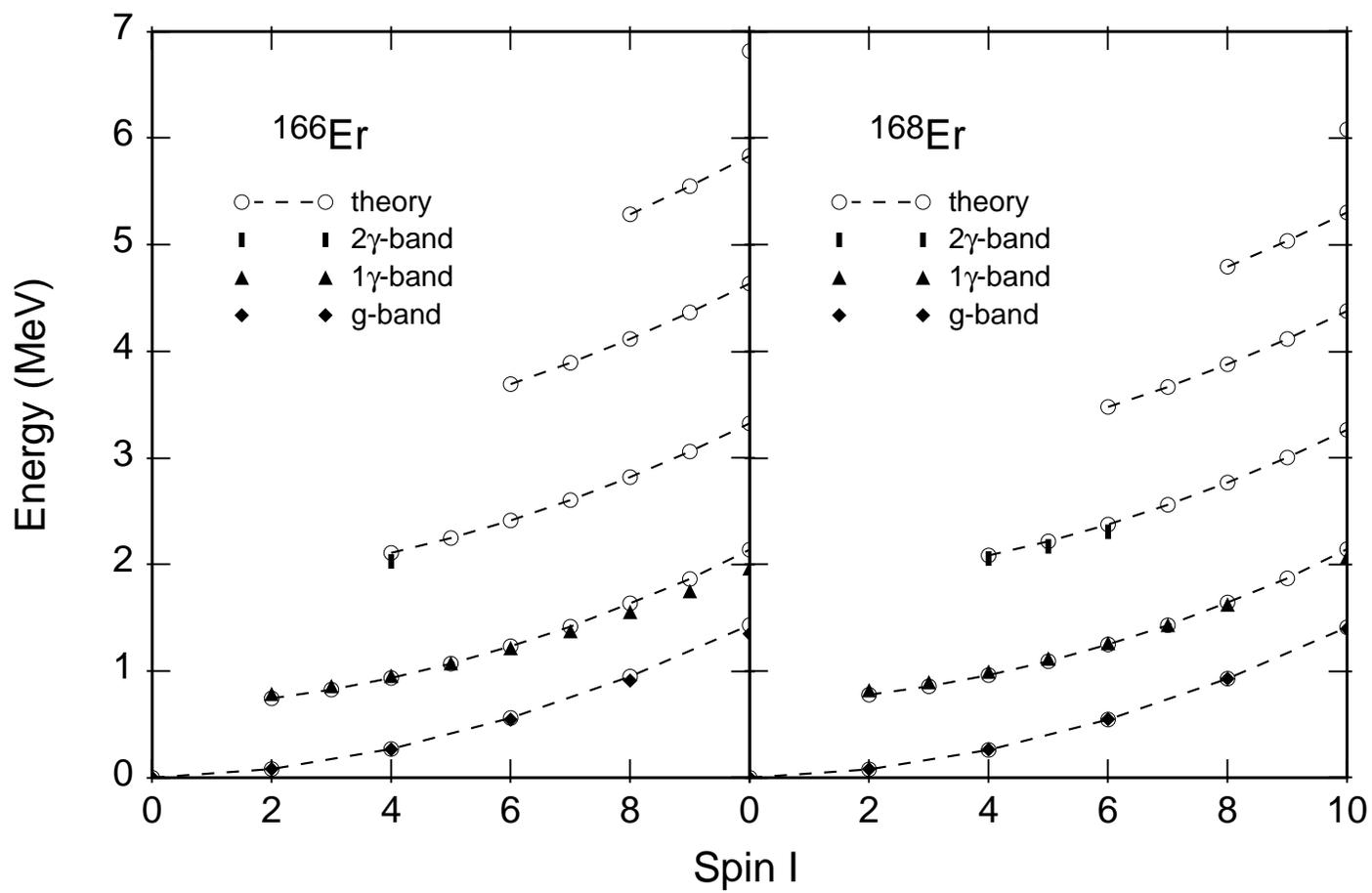
\end{document}